# Optimization Design of Decentralized Control for Complex Decentralized Systems[*]

Ying Huang, Jiyang Dai, Chen Peng

(Key Laboratory of Nondestructive Testing (Nanchang Hangkong University), Ministry of Education, Nanchang 330063, China)

**Abstract:** A new method is developed to deal with the problem that a complex decentralized control system needs to keep centralized control performance. The systematic procedure emphasizes quickly finding the decentralized subcontrollers that matching the closed-loop performance and robustness characteristics of the centralized controller, which is featured by the fact that GA is used to optimize the design of centralized $H_\infty$ controller $\boldsymbol{K}(s)$ and decentralized engine subcontroller $\boldsymbol{K}_\mathrm{T}(s)$, and that only one interface variable needs to satisfy decentralized control system requirement according to the proposed selection principle. The optimization design is motivated by the implementation issues where it is desirable to reduce the time in trial and error process and accurately find the best decentralized subcontrollers. The method is applied to decentralized control system design for a short takeoff and landing fighter. Through comparing the simulation results of decentralized control system with those of the centralized control system, the target of the decentralized control attains the performance and robustness of centralized control is validated.

## 1  INTRODUCTION

Fighters in new generation will incorporate high angle-of-attack maneuver and short take-off and landing capabilities. It has equipped more control surfaces than conventional aerodynamic rudder, such as direct force control and thrust vectoring nozzle. These control surfaces have distributed in the airframe and propulsion systems. How to effectively coordinate these control surfaces to achieve the commands from flight computer and simultaneously reduce the workload of pilot is the most interested problem in integrated flight/propulsion control (IFPC). At present, a centralized approach to IFPC design in [1] is to design a global-level controller to coordinate the airframe and propulsion systems. This approach considers all the interconnections in the aircraft and thus yields an optimal design. However, it results in a high-order controller which is difficult to implement and validate.

\* This work was supported in part by the National Natural Science Foundation of China (61164015) and the Aeronautical Science Foundation of China (2011ZA56021)

One decentralized control approach in [2] is to design airframe and propulsion subsystem controllers in a hierarchical decentralized structure. These subsystem controllers have the lower order and meet the implementation requirements. However, this approach has not considered the interaction between the airframe and propulsion systems. Reference [3] is directly impose an upper triangular structure on the centralized controller and using Riccati-like equations with an iterative process to solve the decentralized controller, however, for high order system problems, can consume significant amounts of computer time. Another decentralized control approach in [4] and [5] is first designing a centralized controller considering the airframe and engine subsystems as one integrated system, and then partitioning the centralized controller into decentralized airframe and propulsion subcontrollers with a specific interconnection structure. During the design, a thumb decision is used to check whether the partitioned controllers can work well and thus leads a very highly iterative design until the integrated control system satisfy the desired performances. If the full envelope of an aircraft is concerned, designers' work will be very large.

In order to handle the decentralized control problem with uncertainty, this paper developed a new approach that put the problem of designing decentralized controllers into optimization problems and solved by using genetic algorithms.

The paper is structured as follows. Section Ⅱ briefly discusses the integrated flight/propulsion model. Section Ⅲ presents the genetic algorithm based centralized $H_\infty$ controller design for the integrated flight/propulsion system as the performance reference of decentralized control. Section Ⅳ presents the optimization design of the decentralized control for the integrated flight/propulsion system. The simulation results and performance analysis are given in Section Ⅴ, and conclusions are given in Section Ⅵ.

## 2  INTEGRATED FLIGHT/PROPULSION MODEL

The vehicle model consists of an integrated airframe and propulsion state-space representation of a modern fighter aircraft powered by a turbofan engine and equipped with 2D thrust vectoring nozzle. The vehicle dynamics is linearized at a flight condition representative of the Short Take-Off and Landing (STOL) approach to landing task (airspeed 61.73 m/s, flight path angle -3°). The model is defined as follows:

$$\dot{x} = Ax + BU, y = Cx + DU \tag{1}$$

where, $A$, $B$, $C$, $D$ are state-space matrices.

The model states $x$, inputs $U$ and outputs $y$ are described below:

$$x = [u, w, q, \theta, h, N_2, N_{25}, P_6, T_{41}]^T$$

$$U = [W_f, A_{78}, A_8, \delta_{tv}]^T$$

$$y = [V, q_v, N_{2P}, R, F_X, F_Z, T_M]^T$$

In the states vector, $u$ is aircraft body axis forward velocity (m/s), $w$ is aircraft body axis vertical velocity (m/s), $q$ is aircraft pitch rate (°/s), $\theta$ is pitch angle (°), $h$ is altitude (m), $N_2$ is engine fan speed (r/min), $N_{25}$ is core compressor speed (r/min), $P_6$ is engine mixing plane pressure (kPa), $T_{41}$ is engine high pressure turbine blade temp (K). In the control inputs vector, $W_f$ is engine main burner fuel flow rate (kg/h), $A_{78}$ is thrust reverser port area (cm$^2$), $A_8$ is main nozzle throat area (cm$^2$), $\delta_{tv}$ is nozzle thrust vectoring angle (°). In the control outputs vector, $V$ is aircraft airspeed (m/s), $q_v=q+0.1\theta$ is pitch variable, which is used in rate command attitude hold (RCAH) control that could track low-frequency pitch rate command and high-frequency pitch angle command, $N_{2P}$ is engine fan speed (percent of maximum allowable r/min at operating condition), $R$ is engine pressure ratio, the last two variables can be the baseline of engine control schedule. $F_X$ and $F_Z$ are total nozzle forces in the x and z direction (N), $T_M$ is total nozzle pitching moment (N·m). In order to reduce condition number of state-space matrices, the model is scaled prior to applying controller design method. The model matrix is full, i.e. the propulsion system states affect the airframe dynamics and vice versa, indicating that there is adequate airframe/propulsion interaction to warrant an integrated control design.

## 3  CENTRALIZED CONTROL OPTIMIZATION DESIGN

Considering the integrated model is high-order complex multivariable model, the GA based mixed sensitivity $H_\infty$ controller design method in [6] is used to provide adequate robustness to modeling uncertainty and model parameter variations with the changing of flight condition and meet the requirements of well tracking and decoupling command abilities. The block diagram of centralized control system for an integrated airframe/propulsion model $G(s)$ is shown in Fig. 1. Note that the control loop allows for feedback of model outputs other than just command tracking error as inputs to the centralized controller $K(s)$. The two degrees of freedom control structure allows simultaneous design of command tracking and loop shaping system. Such control structure is also based on requirements of common flight control system since the overall system requires meeting the desired piloted handling qualities and not just an automatic command tracking system. The rate signal in the model outputs feedback provides for improved damping while position signal feedback improves natural frequency of aircraft mode. Thus this two degrees of freedom control structure is more fit for IFPC than one degree of freedom control structure.

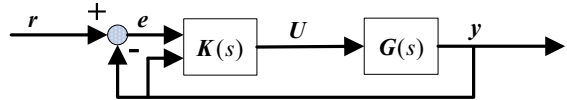

Fig. 1 Centralized control loop

The controller $K$ could be written as $K=[K_e\ K_y]$. The closed-loop transfer function is then given by

$$T_c=[I+G(K_e-K_y)]^{-1}GK_e$$

where the subscript c refers to centralized control system.

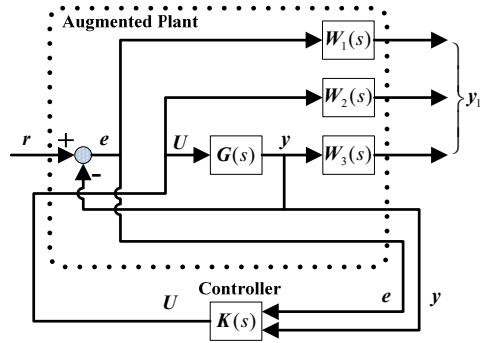

Fig. 2 Block diagram for mixed sensitivity synthesis

The detailed block diagram for the $H_\infty$ mixed sensitivity synthesis is shown in Fig. 2. Define $e=r-y$, $y_1=[W_1(s)e\ W_2(s)U\ W_3(s)y]^T$, $y_2=[e, y]^T$, the matrices $W_1(s)$, $W_2(s)$ and $W_3(s)$ are weighting functions. Then we have

$$\begin{bmatrix} y_1 \\ y_2 \end{bmatrix} = P(s) \begin{bmatrix} r \\ U \end{bmatrix} \tag{2}$$

where $P(s)$ is the augmented plant with the form:

$$P(s) = \left[\begin{array}{c|c} W_1(s) & -W_1(s)G(s) \\ 0 & W_2(s) \\ 0 & W_3(s)G(s) \\ \hline I & -G(s) \\ 0 & G(s) \end{array}\right]$$

We can find that the order of $P(s)$ is equal to one degree of freedom control structure.

The $H_\infty$ mixed sensitivity design problem is to find a controller $K(s)$ that stabilizes the closed-loop system and satisfies

$$\min \|T_{y_1 r}(s)\|_\infty, \text{ or } \|T_{y_1 r}(s)\|_\infty < \gamma, \|T_{y_1 r}(s)\|_\infty = \sup_\omega(\bar{\sigma}[T_{y_1 r}(j\omega)])$$

where $T_{y_1 r}(s)$ is the closed-loop transfer function from the inputs $r(s)$ to the outputs $y_1(s)$.

In the above $H_\infty$ controller design problem, if the weighting functions are chosen and satisfy some constraints, the controller $K(s)$ could be obtained using a modified (by the author) version of the "hinfsyn" function of the Robust Control Toolbox developed by MathWorks Inc. GA-based $H_\infty$ controller design approach in [6] is put the weighting functions as design parameters to tune the controller by optimization such that the design objectives are met. The optimization problem is presented as follows:

Given a nominal plant $G(s)$, to find $(W_1, W_2, W_3)$ such that

$$\min_{W_1, W_2, W_3} \psi_1(W_1, W_2, W_3) = \sum_{i=1}^{n1} k_{1i} r_{1i} \qquad (3)$$

where $\psi_1$ is the fitness function of GA, $r_{1i}$ are the multiple performance indices and $k_{1i}$ are their weights.

Let specifications on the system be of the form

$$\bar{\sigma}(W_1^{-1}(s)) + \bar{\sigma}(W_3^{-1}(s)) \geq 1 \qquad (4)$$

$$\gamma_0(W_1, W_2, W_3) \leq \varepsilon_0 \qquad (5)$$

$$\varepsilon_{i1} \leq \phi_i(W_1, W_2, W_3) \leq \varepsilon_{i2} \qquad (6)$$

where $\gamma_0$ is the $H_\infty$ norm of closed-loop transfer $T_{y_1 r}(s)$, and $\phi_i$ are performance functions representing rise time, overshoot, and bandwidth, etc., and $\varepsilon_0$ and $\varepsilon_i$ are real numbers representing the desired bounds on $\gamma_0$ and $\phi_i$, respectively. Take the bandwidth for instance, if $\phi_i > \varepsilon_{i2}$ or $\phi_i < \varepsilon_{i1}$, then $r_{1i} = |\phi_i - \varepsilon_i|$, else $r_{1i} = 0$.

Based on aircraft handling qualities specifications, and the open-loop analysis of control effectiveness, the inputs and outputs of plant are chosen as $U = [W_f, A_{78}, A_8, \delta_{tv}]^T$ and $y=[V, q_v, N_{2P}, R]^T$ respectively. The desired velocity response represents a well-damped response with minimal overshoot and a rapid settling time with a 90% rise time to a step input in 5 seconds, the desired control bandwidths (here, bandwidth is defined as the frequency at which the response magnitude of the primary commanded variable is -3 dB) are 1 rad/s for $V$ loop, 5 rad/s for $q_v$ loop, 5 rad/s for $N_{2P}$ loop and 10 rad/s for $R$ loop.

The weighting functions $W_1$ and $W_3$ were selected to be first order that to provide adequate frequency response shaping while without increasing excessive order of controller. The $W_2$ was selected as scalar form. Over successive crossover, mutation and selection, the genetic algorithm in [7] drives population of individual toward an optimal solution. In design fitness function of GA, there are some tips to obtain satisfactory solution. For example, when the bandwidth of $q_v$ has achieved 4.9 rad/s, there is a gap of 0.1 rad/s to desired bandwidth 5 rad/s, GA may be stop to optimize this index since 0.1 is only a small proportion in fitness function. We can increase the bound to 5.5 rad/s and last the solution will satisfy desired bandwidth.

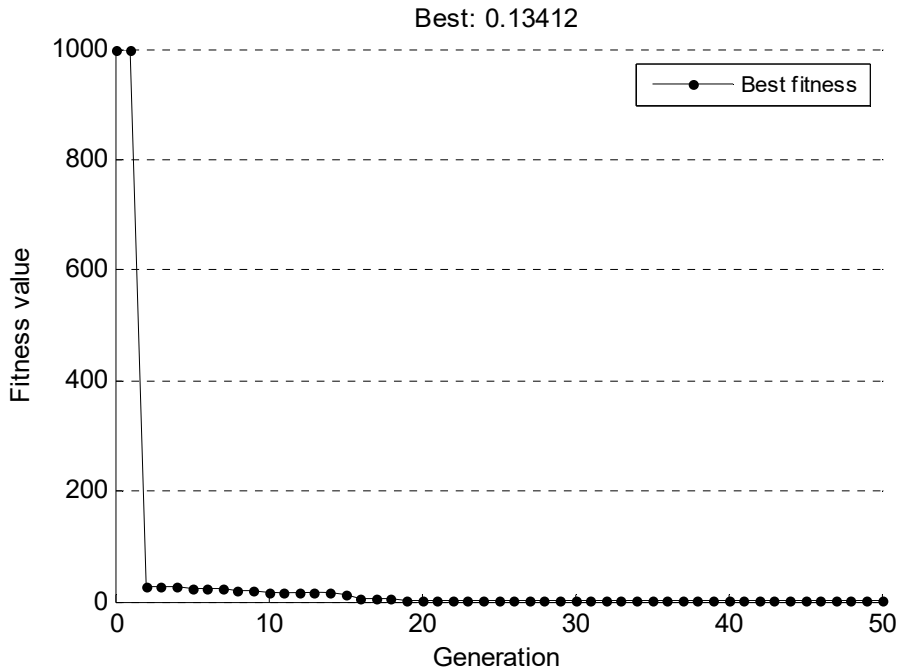

Fig. 3 Best fitness values in centralized control optimization

The augmented plant is 17th order consisting of the 9th order integrated model, first order sensitivity and complementary sensitivity weighting function for the four controlled variables. The optimal weighting functions can be found by GA optimization, and simultaneously the 17th order centralized controller can be obtained. We have checked the controller itself is stable. Reference [8] has mentioned the significance of controller stability. Further the stable controller is beneficial for controller reduction that will be used in decentralized control. The best fitness values during the optimization design of centralized control are shown in Fig. 3, although first two generation fitness values are very high due to initial random individuals, finally it is convergent to global optimal value. The centralized control closed-loop system performance is evaluated by Bode diagram and step command response. An example result is shown in Fig. 4 in terms of closed-loop velocity frequency response to all the commanded inputs. We can find that closed-loop system could track velocity commands to the desired bandwidth with small response in velocity to other commands. By analysis, the command tracking bandwidths of centralized closed-loop system are 1.47, 5.79, 5.95 and 10.89 rad/s respectively, which meet the handling qualities requirements well. For compared with decentralized control system, the time-domain response of closed-loop system will given in the later section.

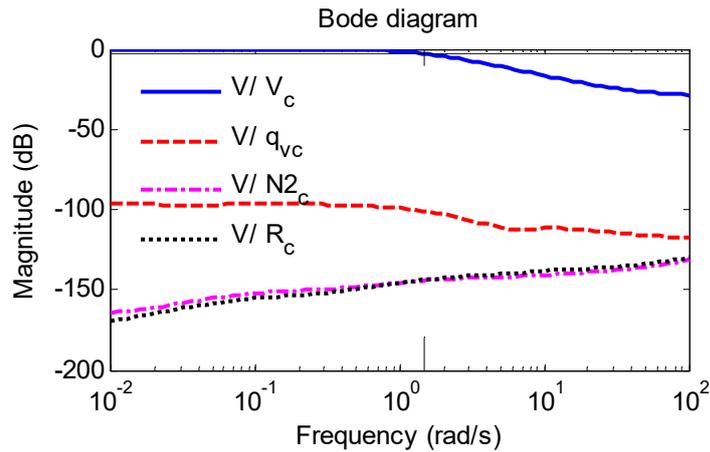

Fig. 4 Closed-loop Bode plot of velocity response to command inputs

## 4 DECENTRALIZED CONTROL OPTIMIZATION DESIGN

### 4.1 DECENTRALIZED CONTROL PROBLEM

The desired structure of the decentralized control will depend on the coupling between the various subsystems and on practical considerations related to integration of the independently controlled subsystems. As pointed out in [5], the most suitable control structure for the IFPC problem is hierarchical with the airframe (flight) controller generating commands for the aerodynamic control surfaces as well as for the propulsion subsystem. This decentralized, hierarchical control structure is shown in Fig. 5 where the subscripts and superscripts *a* and *e* refer to airframe and propulsion system (engine) quantities, respectively, *r* refers to commands, and the variables *y* are the controlled outputs of interest with *e* being the corresponding errors. The intermediate variables $y_{ea}$ represent propulsion system quantities that affect the airframe, for example propulsion system generated forces and moments.

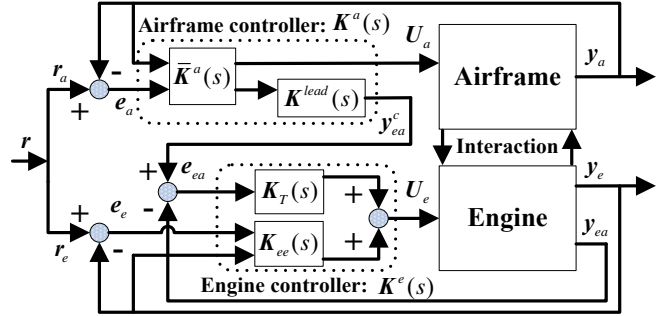

Fig. 5 Decentralized control loop

The decentralized control problem of Fig. 5 can be stated as follows:

Given a centralized controller $K(s)$ satisfying

$$U = K(s)\begin{bmatrix} e \\ y \end{bmatrix}, U = \begin{bmatrix} U_a \\ U_e \end{bmatrix}, e = \begin{bmatrix} e_a \\ e_e \end{bmatrix}, y = \begin{bmatrix} y_a \\ y_e \end{bmatrix}$$

where $U_a$ and $U_e$ refer to airframe and engine control inputs, and $e_a$ and $e_e$ refer to airframe and engine commands tracking errors, respectively.

To find decentralized airframe and engine subcontrollers $K^a(s)$ and $K^e(s)$, respectively, with:

$$\begin{bmatrix} U_a \\ y_{ea}^c \end{bmatrix} = K^a(s)\begin{bmatrix} e_a \\ y_a \end{bmatrix}, U_e = K^e(s)\begin{bmatrix} e_e \\ y_e \\ e_{ea} \end{bmatrix}$$

such that the closed-loop performance and robustness with the subcontrollers $K^a(s)$ and $K^e(s)$ match those with the centralized controller $K(s)$ to a desired accuracy. Furthermore, the engine subcontroller $K^e(s)$ could tracking interface variable commands and allow for independent verify the propulsion system.

## 4.2 DECENTRALIZED CONTROLLER DESIGN

Let the controlled plant $G'(s)$ be of the form

$$\begin{bmatrix} y_a \\ y_e \\ y_{ea} \end{bmatrix} = \begin{bmatrix} G_{aa}(s) & G_{ae}(s) \\ G_{ea}(s) & G_{ee}(s) \\ G_{ea}^a(s) & G_{ea}^e(s) \end{bmatrix} \begin{bmatrix} U_a \\ U_e \end{bmatrix} = G'(s) \begin{bmatrix} U_a \\ U_e \end{bmatrix} \qquad (7)$$

In the decentralized control loop, the controlled plant $G'(s)$ has more interface variables than $G(s)$ in the centralized control loop. Thus the superscript apostrophe is being added in the $G'(s)$.

## 4.2.1 ASSIGNMENT OF PLANT'S INPUTS AND OUTPUTS

（Ⅰ）Assign the plant inputs and outputs to be either engine, airframe or interface variables. Through open-loop analysis indicated that thrust vectoring nozzle $\delta_v$ is primarily a pitch effector, thus $U_a=[\delta_v]$, $y_a=[V, q_v]^T$. The main fuel flow rate $W_f$, thrust reverser port area $A_{78}$, and main nozzle thrust area $A_8$ mainly affect engine's operation, thus $U_e=[W_f, A_{78}, A_8]^T$, $y_e=[N_{2P}, R]^T$.

（Ⅱ）Choose the interface variables. Generally, the interaction from engine to airframe is generated by forces and moments. Choose the forces and moments as the interface variables could control the coupling of engine to airframe. For simplify the tracking controller, we have made a more detailed investigation. In the centralized control loop, we add thrusts $F_X$, $F_Z$ and pitch moment $T_M$ as the model outputs without feedback to controller to observe the responses to step airframe commands. The interface variables and thrust nozzle angle closed-loop response to step velocity command with the centralized controller are shown in Fig. 6. Note that the quantities in the figure are scaled quantities that the response magnitudes could be directly compared. The results reflect that the changes of $F_Z$ and $T_M$ are mainly affected by thrust nozzle angle $\delta_v$ while $F_X$ is to track the step velocity command. Thus, $F_X$ is the most critical variable which need to control. Step pitch variable command will have the same conclusion. We remove $F_Z$ and $T_M$, and only retain $F_X$ as the interface variable. This simplification also reduces the complexity of engine subsystem controller. In the last, the simulation results of decentralized closed-loop control will reflect the reasonability for only retaining $F_X$.

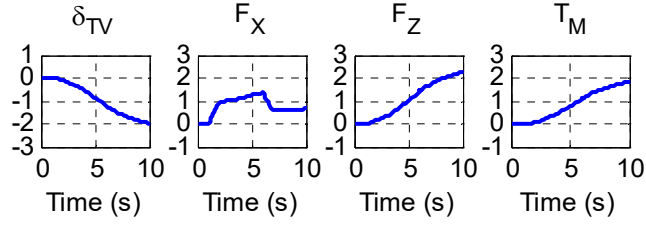

Fig. 6 Interface variables and thrust nozzle angle response to step velocity command

## 4.2.2 AIRFRAME SUBCONTROLLER ($\bar{K}^a$ AND $K^{lead}$) DESIGN

(Ⅰ) **Subcontroller** $\bar{K}^a$ With the centralized controller $K(s)$ in series with the engine subsystem loop and $r_e$ set as zero, as shown in Fig. 7.

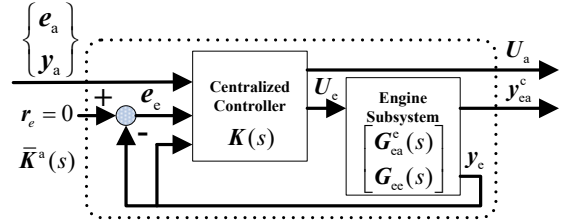

Fig. 7 Configuration of airframe subcontroller

This connection is defined as the flight controller $\bar{K}^a$ describing a transfer function from $[e_a, y_a]^T$ to $[U_a, y_{ea}^c]^T$, whose state-space representation can be obtained using algebraic manipulation. The order of $\bar{K}^a$ is 21st order that equal to the sum of the orders of centralized controller and engine subsystem (4th order). Using the improved balanced realization reduction approach in [9], $\bar{K}^a$ can be reduced to an 8th order approximation of the 21st order. We checked $\bar{K}^a$ and found that is stable. Note that we didn't reduce the order of centralized controller beforehand for keep the characteristics of it in the $\bar{K}^a$. The singular values of full and reduced order flight controller are compared in Fig. 8 and indicate that there is little change for controller reduction.

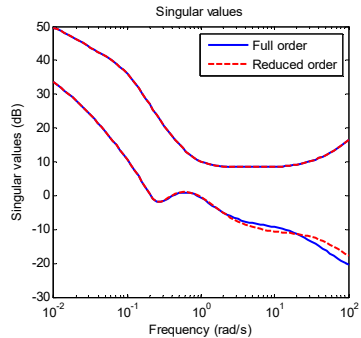

Fig. 8 Singular values of full and reduced order flight controller

(Ⅱ) **Lead filter** $K^{lead}$ is designed to be a lead filter to compensate for the delay of engine subsystem tracking $y_{ea}^c$ command due to the process of engine combustion. In general, there will be a design trade-off based on practical considerations. High lead compensation can result in saturation of the engine actuators, whereas low lead compensation will require large $y_{ea}^c$ tracking bandwidth. Since controller $K_T$ provides decoupled tracking of $y_{ea}^c$, $K^{lead}$ can simply be of the form

$$K^{lead}(s) = diag\left[\frac{s+a}{a} \quad \frac{b}{s+b}\right], a < b \tag{8}$$

Here, the lead compensation was chosen to be a=10, b=30, which resulting in an effective bandwidth of 30 rad/s for interface variable.

The airframe subcontroller $K^a(s)$ consists of $\bar{K}^a$ and $K^{lead}$ according to the structure in Fig. 5.

### 4.2.3 ENGINE SUBCONTROLLER ($K_T$ AND $K_{ee}$) DESIGN

(Ⅰ) **Interface variable tracking controller** $K_T$ is designed to track $y_{ea}^c$ commands and provides decoupling between $r_e$ and $y_{ea}^c$ responses. The control loop to determine $K_T$ is shown in Fig. 9. Noticed that the control problem is non-square where control loop has three control inputs and only one controlled output. For improve $H_\infty$ control problem solving efficiency, we add variables $y_e$ as the engine outputs. After the controller $K_T$ is solved, then remove $y_e$ and only retain $y_{ea}$. The optimization procedure to solve the $K_T$ will discuss in the next section.

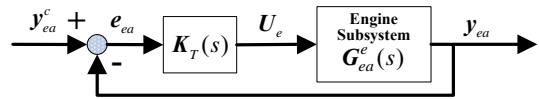

Fig. 9 Control loop to determine interface variable tracking controller

(Ⅱ) The $K_{ee}$ part of the engine subcontroller is regarded as a reduced-order approximation of the $K_{ee}(s)$ block of the centralized controller. In order to reduce subcontroller complexity, it is important to keep the order of $K_{ee}$ as low as possible while keep the input/output characteristics of the corresponding centralized controller block $K_{ee}(s)$. The centralized controller can be partitioned as

$$K(s) = \begin{bmatrix} K_{aa}(s) & K_{ae}(s) \\ K_{ea}(s) & K_{ee}(s) \end{bmatrix}$$

$K_{ee}$, which is abstracted from $K(s)$, can be reduced from 17th order to a 10th order

controller by using the balanced realization controller reduction approach. The singular values of full and reduced order engine controller are compared in Fig. 10 and indicate that there is little change for controller reduction.

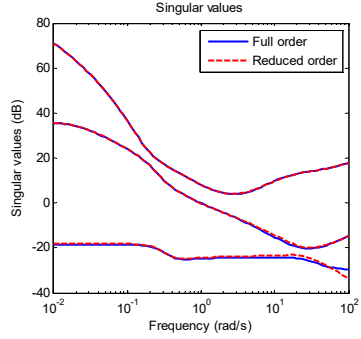

Fig. 10 Singular values of full and reduced order engine controller

The engine subcontroller $K^e(s)$ consists of $K_T$ and $K_{ee}$ according to the structure in Fig. 5. The state-space realization for decentralized control closed-loop transfer function $T_d$ could be obtained according to input/output connections of various blocks shown in Fig. 5. The subscript d refers to decentralized control.

## 4.3 THE DECENTRALIZED CONTROL OPTIMIZATION PROBLEM

From the decentralized subcontroller design procedure, we find a relation between centralized control and decentralized control. The airframe subcontroller $K^a(s)$ is combined by centralized controller and engine subsystem. The engine controller $K_{ee}$ is abstract from centralized controller. Once the centralized controller is properly designed, both of $K^a(s)$ and $K_{ee}$ are fixed. The controller $K_T$ is the only one part that can be tuned in the decentralized control closed-loop transfer function $T_d$. So we investigate the controller optimization method that can preserve the closed-loop stability and minimize the performance degradation of decentralized control closed-loop system.

The idea of controller reduction with stability criteria in [10] is used that the closed-loop stability is guaranteed and the closed-loop performance degradation is limited if $\|T_d-T_c\|_\infty$ is sufficiently small. Simultaneously, the closed-loop transfer function $T_d$ does not necessarily make the error $\bar{\sigma}(T_d-T_c)(j\omega)$ sufficiently small uniformly over all frequencies. The approximation error only has to be made small over those critical frequency ranges that affect the closed-loop stability and performance.

Without loss of generality, the decentralized control closed-loop system is stable if:
$$\|T_d - T_c\|_\infty < 1 \tag{9}$$

where $\|T_d\text{-}T_c\|_\infty$ denotes the maximum value of $\bar{\sigma}(T_d\text{-}T_c)(j\omega)$ over all frequencies.

The decentralized controller design problem is converted into optimization problem that minimize the error $\bar{\sigma}(T_d\text{-}T_c)(j\omega)$ by select $K_T$, i.e.,
$$\min_{K_T} \|T_d(j\omega) - T_c(j\omega)\|_\infty \tag{10}$$

Consider the $H_\infty$ norm is numerically difficult to calculate in the optimization process, the error can be expressed as the following form:
$$\min_{W_S, W_C, W_T} \psi_2(W_S, W_C, W_T) = \frac{1}{m}\sum_{1}^{m}(\bar{\sigma}[T_d(j\omega) - T_c(j\omega)])^2 \tag{11}$$

where $m$ is the number of frequency points in the interested frequency region $(\omega_1, \omega_2)$, $\bar{\sigma}$ is the maximum singular value at the frequency point $\omega$.

Simultaneously, the $K_T$ should meet the specifications of tracking interface variable, such as tracking bandwidth, rise time and overshoot. Previous experience has shown that the higher bandwidth of $K_T$ the smaller error for closed-loop system. Thus the bandwidth of $K_T$ is chosen to be the bandwidth of engine actuator, is 30 rad/s. Like the fitness function of centralized controller, for the 4th order engine subsystem as shown in Fig. 9, the sensitivity weight $W_S$, the complementary weight $W_T$, and the control weight $W_C$ are as design parameters, GA is used to search the optimal weighting functions and thus to obtain the controller $K_T$. The fitness function of GA has the representation:
$$f_2 = k_\psi \psi_2 + \sum_{j=1}^{n2} k_{2j} r_{2j} \tag{12}$$

where $\psi_2$ and $r_{2j}$ are the performance indices and $k_\psi$ and $k_{2j}$ are their weights.

The main frequency band of aircraft is $(\omega_1, \omega_2) = (0.01, 100)$(rad/s) with 20 frequency points per decade resulting in 81 points. The best fitness values of GA optimization are shown in Fig. 11. GA stopped at 46th generation due to the change in the fitness value is less than preallocated tolerance. The resulting 10th order controller $K_T$ is obtained and reduced to 4th order using balanced realization. The full and reduced order controller are stable and their singular values are compared in Fig. 12 and indicate that there is almost no change. For compare the effects of controller reduction, the full order airframe and engine subcontrollers are also studied in the optimization design. The maximum singular value of error $\bar{\sigma}(T_d\text{-}T_c)(j\omega)$ with full and reduced order two cases are shown in Fig. 13. For the full order decentralized control closed-loop system, $\|T_{\text{fd}}-T_{\text{c}}\|_\infty=0.1$, and $\|T_{\text{rd}}-T_{\text{c}}\|_\infty=0.2$ for the reduced order decentralized control closed-loop system, thus ensuring stability condition (9) and also minimized the error with the centralized control closed-loop system.

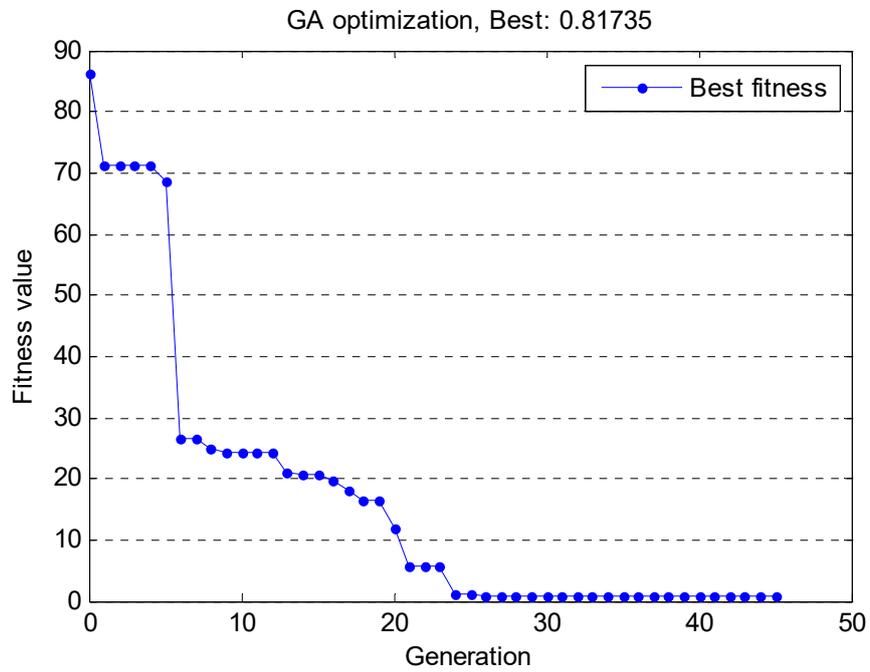

Fig. 11 Best fitness values in reduced order decentralized control optimization

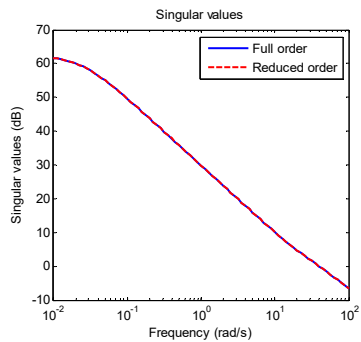

Fig. 12 Singular values of full and reduced order interface variable controller

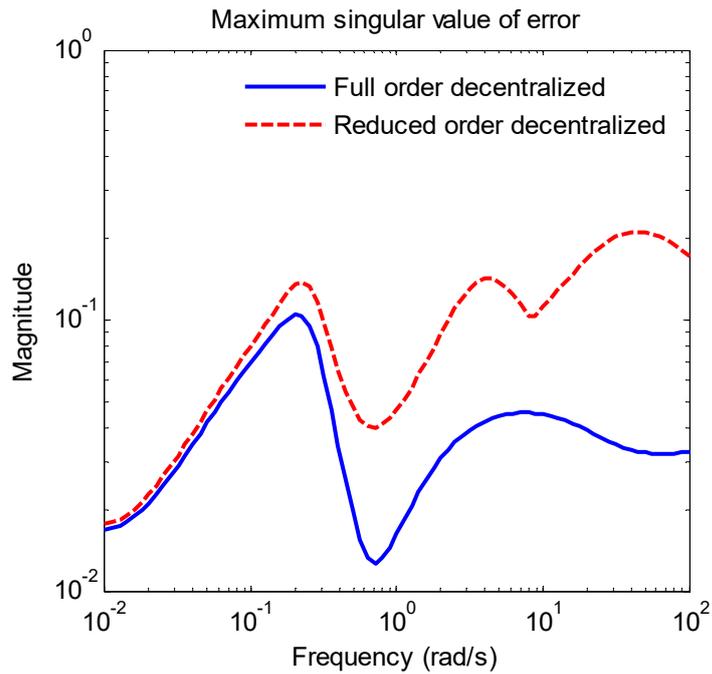

Fig. 13 Maximum singular value of error

## 5 SIMULATION AND ANALYSIS

Comparisons are performed between the closed-loop responses with the centralized controller, full order and reduced order decentralized subcontrollers. An example comparison for step velocity command $V_c$ is shown in Fig. 14. The responses in Fig. 14 indicate that the full and reduced order decentralized subcontrollers maintain the performance of tracking velocity command without overshoot and decoupling of pitch variable, engine fan speed, and engine pressure ratio by the centralized controller.

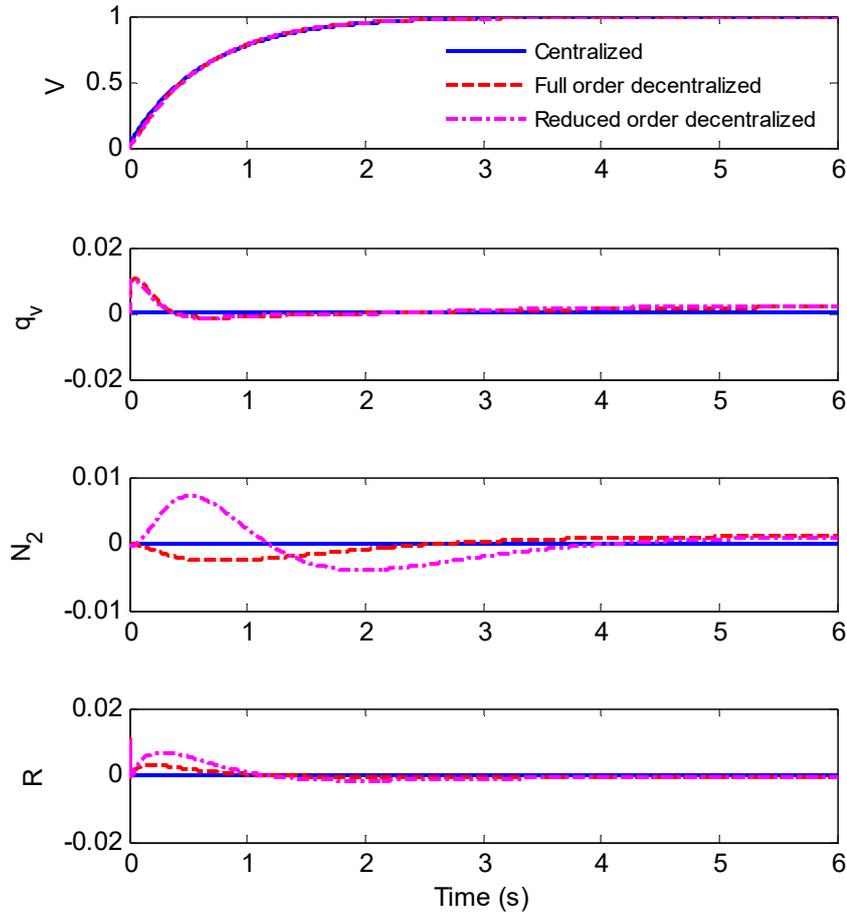

Fig. 14 Outputs responses to a velocity command for centralized, full and reduced order decentralized control

The results presented so far have focused on comparing the performance achieved with the decentralized subcontrollers to that achieved with the centralized controller. Robustness issues are also of importance in practical control design. Robustness analysis was performed using structured singular values $\mu$ for gain and phase variations occurring at the controlled outputs and the results are shown in Fig. 15 for the centralized, full and reduced order decentralized control closed-loop systems. The calculation for robustness analysis using structured singular values is referenced in [11]. From Fig. 15, the maximum value over frequency of the structured singular value, is 1.1226 with the centralized controller, 1.1226 with the full order decentralized subcontrollers and 1.1082 with the reduced order decentralized subcontrollers, and their stability margins SM=1/$\mu_{max}$ are 0.8908, 0.8907 and 0.9024 respectively. The multivariable gain margins and phase margins are given by:

$$GM = 20 \times \log_{10}\left(\frac{1}{1 \pm SM}\right), PM = \pm 2\sin^{-1}(SM/2)$$

The final gain margins are -5.5323～19.2249 dB, -5.5323～19.2249 dB and -5.586～20.2094 dB for the centralized, full and reduced order decentralized control closed-loop system, respectively, and similarly phase margins of ±52.901°, ±52.893° and ±53.644°, respectively, for simultaneous gain or phase variations occurring in all the loops at the controlled outputs. Thus optimization design approach for decentralized control can maintain the robustness characteristics of centralized control.

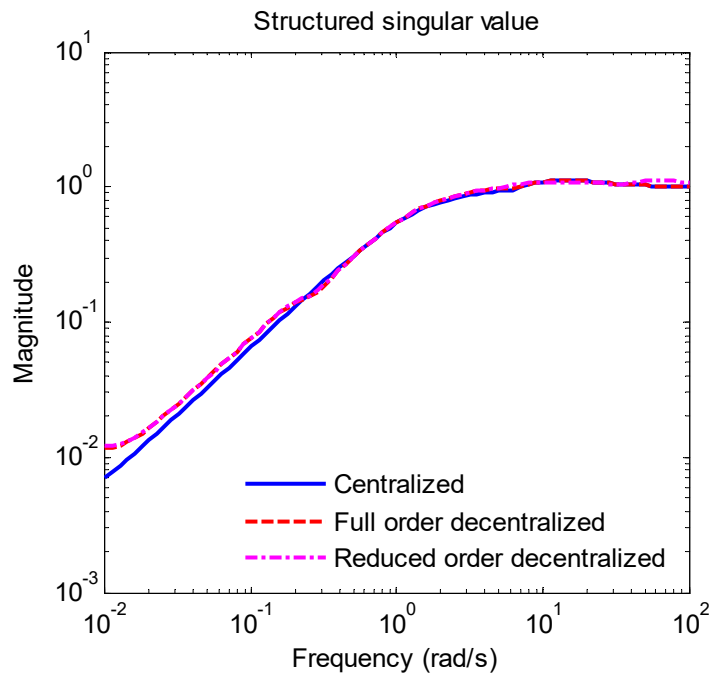

Fig. 15 Structured singular value for robustness analysis

The results of simulation and robustness analysis indicate that the optimal controller can be efficiently obtained through optimization. In the decentralized control, the order of subcontrollers can be reduced to the lowest as long as their performance are satisfactory, and the optimization design approach proposed in this paper is still effective for maintain the performance and robustness of centralized control.

# 6  CONCLUSIONS

A systematic procedure is presented for designing decentralized integrated flight propulsion control (IFPC) system. The procedure emphasizes efficiently finding the decentralized subcontrollers that matching the closed-loop performance and robustness characteristics of the centralized controller, which is featured by the fact that GA is used to optimize the design of centralized $H_\infty$ controller $K(s)$ and decentralized engine subcontroller $K_\text{T}$, and that only one interface variable needs to satisfy decentralized control system requirement according to the proposed selection principle. The optimization design is motivated by the implementation issues where it is desirable to reduce the time in trial and error process and accurately find the best decentralized subcontrollers. The optimization design procedure were described and demonstrated through application to IFPC design for a short take-off and vertical landing aircraft. The simulation results have shown that the optimization design procedure resulted in low order airframe and engine subcontrollers that maintain the performance and robustness achieved by the centralized controller.